\documentclass[fleqn,usenatbib,twocolumn]{mnras}

\usepackage{newtxtext,newtxmath}

\usepackage[T1]{fontenc}
\usepackage{ae,aecompl}

\usepackage{graphicx}                           
\usepackage{amsmath}                            
\usepackage{collcell}                           
\usepackage{float}                              
\usepackage{flafter}                            
\usepackage{booktabs}                           
\usepackage{threeparttable}                     

\newcommand{\dmunits}{\,pc\,cm$^{-3}$\:}                
\newcommand{\funits}{\,Jy\,ms\:}                        
\newcommand{\JperHz}{\,J\,Hz$^{-1}$\:}                  
\newcommand{\uclu}[2]{\,$_{-#1}^{+#2}$\:}               
\newcommand{\EMin}{\,$E_{\mathrm{min}}$\:}              
\newcommand{\EMax}{\,$E_{\mathrm{max}}$\:}              
\newcommand{\DMIGM}{$\mathrm{DM}_{\mathrm{IGM}}$\:}     
\newcommand{\DMHOST}{$\mathrm{DM}_{\mathrm{Host}}$\:}   
\newcommand{\FRB}[1]{FRB\,#1}                           
\newcommand{\PlotSize}{0.85}                            

\newcommand{\includepic}[1]{\includegraphics[width=6.0cm,height=5.0cm,keepaspectratio=true]{#1}}
\newcolumntype{i}{@{\hspace{1ex}}>{\collectcell\includepic}c<{\endcollectcell}}

\defcitealias{Macquart_Ekers_2018_2}{\,M\&E\:2018 }
\newcommand{\ME}{\citetalias{Macquart_Ekers_2018_2}}

\defcitealias{Arcus_etal_2020}{Paper~I}
\newcommand{\PaperI}{\citetalias{Arcus_etal_2020}}

\title[Parkes and FAST FRB DM Distribution]{Comparison of the Parkes and FAST FRB DM Distribution}

\author[Arcus et al.]{
    W. R. Arcus,$^{1}$\thanks{E-mail: wayne.arcus@icrar.org (WA)},
    C. W. James$^{1}$, R. D. Ekers$^{1,2}$, R. B. Wayth$^{1}$.
    \\
    $^{1}$ International Centre for Radio Astronomy Research, Curtin University, GPO Box U1987, Perth, WA 6845, Australia\\
    $^{2}$ CSIRO Astronomy and Space Science, P.O. Box 76, Epping, NSW 1710, Australia\\
}

\date{Accepted XXX. Received YYY; in original form ZZZ}

\pubyear{2022}

\begin{document}
\label{firstpage}
\pagerange{\pageref{firstpage}--\pageref{lastpage}}
\maketitle


\begin{abstract}
    We model the Fast Radio Burst (FRB) dispersion measure (DM) distribution for the Five-hundred-meter Aperture Spherical Telescope (FAST) and compare this with the four FRBs published in the literature to date. We compare the DM distribution of Parkes and FAST, taking advantage of the similarity between their multibeam receivers. Notwithstanding the limited sample size, we observe a paucity of events at low DM for all evolutionary models considered, resulting in a sharp rise in the observed cumulative distribution function (CDF) in the region of $1000\lesssim\mathrm{DM}\lesssim2000$\,pc\,cm$^{-3}$. These traits could be due to statistical fluctuations ($0.12 \le p \le 0.22$), a complicated energy distribution or break in an energy distribution power law, spatial clustering, observational bias or outliers in the sample (e.g., an excessive \DMHOST as recently found for \FRB{20190520B}). The energy distribution in this regime is unlikely to be adequately constrained until further events are detected. Modelling suggests that FAST may be well placed to discriminate between redshift evolutionary models and to probe the helium ionisation signal of the IGM.
\end{abstract}


\section{Introduction}
The FRB population has been established as being cosmological in nature, with dispersion measures (DMs) typically comprising dominant contributions attributable to propagation through the intergalactic medium (IGM). Under such circumstances, it is viable to use the measured FRB DM as a proxy for redshift \citep[see][and references therein]{Macquart_etal_2020}.

In a previous work \citep[][hereinafter \PaperI]{Arcus_etal_2020}, we utilised the \ME model of \citet{Macquart_Ekers_2018_2} to compare the FRB DM distributions, determined from Parkes and ASKAP radio telescope samples, to infer the common FRB population parameters of fluence spectral index, $\alpha$, and the energy function slope, $\gamma$, assuming a power law energy function. Therein we further assumed the telescopes observed the same FRB population and made use of the fact that the telescopes had different survey fluence limits, $F_{0}$ \citep{Arcus_etal_2020}. We found: i) no evidence that the FRB population evolves faster than linearly with respect to the cosmic star formation rate (CSFR) for the population sample analysed; ii) the spectral index and energy curve slope were degenerate; and iii) the modelled ASKAP and Parkes DM distributions were consistent with a range of source evolution models.

In this work, we utilise, \textit{a priori}, these same fitted parameters to model the DM distribution predicted to be observed by FAST -- the Five-hundred-meter Aperture Spherical Telescope \citep[see, e.g.,][]{Li_etal_2018,Zhu_etal_2020,Niu_etal_2021}. We admit FAST samples in order to introduce higher sensitivity readings, thereby probing deeper into the Universe and enhancing observational discrimination. Moreover, given that Parkes and FAST utilise a nearly identical multibeam receiver (19 versus 13 beams), this permits a similar analysis to be undertaken in order to reduce systematic errors.

We extend the approach described in \PaperI\: by accounting for the telescope beam pattern, generalising the fitted sensitivity relation of the telescope's back-end and by accounting for temporal smearing following the approach of \citet{Cordes_McLaughlin_2003}.

Furthermore, we examine the FRB DM distribution predicted to be detected by FAST, and compare these results with the published FAST FRB event data, noting that, hitherto, only a very limited FRB event sample is available.

In \S \ref{sec:DMDistribution}, we describe modifications made to the DM distribution model and we compare the resultant distributions with published FRB events. In \S \ref{sec:Discussion} and \S \ref{sec:Conclusion} we discuss, respectively, the implications of our results (which may have relevance to SKA1-mid) and our conclusions.

\section{The DM Distribution}
\label{sec:DMDistribution}
The \ME model is principally based on relating the DM distribution, $dR_{F}/d\mathrm{DM}$, to the redshift distribution, $dR_{F}/dz$, for fluences above a survey fluence limit, $F_{0}$, via eq.(\ref{eqn:RedshiftDistribution}) and eq.(\ref{eqn:DMDistribution}), and assuming that redshift, $z$, and $\mathrm{DM}$ are bijective for a homogeneous IGM \citep[i.e., one-to-one hence $z \cong \mathrm{DM}$,][]{Macquart_Ekers_2018_2}. Here

\begin{equation}
    \begin{split}
    \dfrac{dR_{F}}{dz}&(F_{\nu} > F_{0},z) = 4 \pi D_H^5\left(\dfrac{D_M}{D_H}\right)^4 \dfrac{(1+z)^{\alpha-1}}{E(z)} \psi_{n}(z) \\
    &\dfrac{(1+z)^{2-\alpha}}{4 \pi D_{L}^{2}(z)}
    \begin{cases}
        0 & F_{0} > F_{\mathrm{max}} \\
        \left(\dfrac{F_{\mathrm{max}}^{1-\gamma} - F_0^{1-\gamma}}{F_{\mathrm{max}}^{1-\gamma} - F_{\mathrm{min}}^{1-\gamma}}\right) & F_{\mathrm{min}} \le F_{0} \le F_{\mathrm{max}} \\
        1 & F_{0} < F_{\mathrm{min}}
    \end{cases}
    \end{split}
    \label{eqn:RedshiftDistribution}
\end{equation}
\\

\noindent and

\begin{equation}
    \dfrac{dR_F}{d\mathrm{DM}}(F_{\nu} > F_{0},z) = \dfrac{dR_F}{dz}(F_{\nu} > F_0, z)/\dfrac{d\overline{\mathrm{DM}}}{dz}(z)\mathrm{,}
\label{eqn:DMDistribution}
\end{equation}
\\

\noindent where $R_{F}$ is the (fluence) differential FRB event rate in the observer's frame of reference per unit solid angle. Other symbol definitions relevant to the model are listed in Table \ref{tab:ListOfSymbols}\footnote{Reproduced from \citet{Arcus_etal_2020} and provided for convenience.}.

Here, the CSFR, $\psi_{n}(z)$, has been adopted from \citet{Madau_Dickinson_2014} and the mean DM of an homogeneous IGM, $\overline{\mathrm{DM}}(z)$, from \citep{Ioka_2003}; they are respectively given by equations eq.(\ref{eqn:CSFR}) and eq.(\ref{eqn:DM})\footnote{We equate $\overline{\mathrm{DM}} = \mathrm{DM}_{\mathrm{IGM}}$ in the DM budget (i.e., $\mathrm{DM}_{\mathrm{Obs}} = \mathrm{DM}_{\mathrm{MW}} + \mathrm{DM}_{\mathrm{Halo}} + \mathrm{DM}_{\mathrm{IGM}} + \mathrm{DM}_{\mathrm{Host}}/(1+z)$) and assume a constant host contribution \citep[see eq.(1) of][and details therein]{Arcus_etal_2020}.} thus

\begin{equation}
    \psi_{n}(z) = K \left(\dfrac{0.015 (1 + z)^{2.7}}{1 + ((1 + z)/2.9)^{5.6}}\right)^{n} \mathrm{yr^{-1} Mpc^{-3}}
    \label{eqn:CSFR}
\end{equation}
\\

\noindent and

\begin{equation}
    \begin{split}
    \overline{\mathrm{DM}}(z)=&\dfrac{3H_0c\Omega_b}{8 \pi G m_p}\\
    &\int_{0}^{z} \dfrac{(1+z') \left[\frac{3}{4} X_{\mathrm{e,H}}(z') + \frac{1}{8}X_{\mathrm{e,He}}(z')\right] }{\sqrt{(1+z')^3 \Omega_m + \Omega_{\Lambda}}} dz'\text{.}
    \end{split}
    \label{eqn:DM}
\end{equation}
\\

\noindent The index, $n$, of eq.(\ref{eqn:CSFR}) relates to three redshift evolutionary scenarios: $n=0$ -- none; $n=1$ -- linear; and $n=2$ -- quadratic evolution with respect to the CSFR.

Consistent with \citet{Arcus_etal_2020} and throughout this work, we take the ionised fraction of Hydrogen and Helium to be $X_{\mathrm{e,H}}=1$ for $z<8$ and $X_{\mathrm{e,He}}=1$ for $z<2.5$ respectively, or zero otherwise. We take $\alpha$ to refer to the fluence spectral index, defined such that $F_{\nu} \propto \nu^{-\alpha}$, and we assume a $\Lambda$CDM universe consistent with the \citet{Plank_Results_2013} -- i.e., $(h,\: H_{0},\: \Omega_b,\: \Omega_m,\: \Omega_{\Lambda},\: \Omega_k) = (0.7,\: 100\:h\,\mathrm{km s}^{-1} \mathrm{Mpc}^{-1},\: 0.049,\: 0.318,\: 0.682,\: 0)$.

\subsection{Model Extensions}
To better account for instrumental effects, we extend the model by: i) generalising the sensitivity formulation using the method introduced by \citet{Cordes_McLaughlin_2003}, to allow for ready generalisation to different telescopes; and ii) incorporating the effect of the beam function on fluence limit, hence on the population probed \citep[see, e.g.,][]{Macquart_Ekers_2018_2,Connor_2019,Luo_etal_2020,James_etal_2021_1}.

 \subsubsection{Instrument Response}
Since the limiting fluence of a survey is a function of DM, hence redshift \citep{Arcus_etal_2020}, we make the substitution given by eq.(\ref{eqn:F0Substitution}), by including the telescope beam function, $B(\theta)$. We normalise the mean search sensitivity, $\overline{\eta}(z)$, by interpreting $F_{0}$ as $F_{0}(z=0)$. The sensitivity at $\mathrm{DM}=0$ is not important in this context as we are principally concerned with comparing relative event rates via the DM distribution curve shape. (The mean search sensitivity accounts for the mean loss of sensitivity due to DM smearing of the FRB at redshift, $z$. Furthermore, and throughout this work, we take DM to be synonymous with the \DMIGM, determined from the DM budget unless specifically noted otherwise.)

\begin{equation}
    F_{0}(z) \rightarrow F_{0}(z,\theta) = \left\{F_{0}/(\overline{\eta}(z) B(\theta)) : \overline{\eta}(z=0)=1\right\}\text{.}
\label{eqn:F0Substitution}
\end{equation}\\

Integrating eq.(\ref{eqn:DMDistribution}) over the telescope's Field of View (FoV) yields

\begin{equation}
    \begin{split}
    \dfrac{dR_F}{d\mathrm{DM}}(F_{\nu}>F_0(z,\theta))=&\\
    \int_{0}^{2 \pi}\int_{0}^{\theta_{b}}\dfrac{dR_F}{dz}/&\dfrac{d\overline{\mathrm{DM}}}{dz}\cdot \sin \theta d\theta d{\phi}
    \mathrm{,}
    \end{split}
\label{eqn:DMDistBeamCorrected}
\end{equation}
\\

\noindent where $\theta_{b}$ is the applicable beam integration limit.

For a detected burst of width, $w$, the detection sensitivity varies as $w^{-1/2}$ \citep{Cordes_McLaughlin_2003}, resulting from the burst energy being spread over time. Here, we are primarily interested in DM-dependent effects, introduced by different spectral and temporal resolutions used for incoherent de-dispersion searches. The effects of scattering are included in the distribution of FRB widths modelled in \PaperI, so that only differences in FRB widths (e.g. due to scattering) between Parkes and FAST will skew our model. There is, however, no evidence for a correlation between DM and scattering \citep[see, e.g.,][\S7.3]{ChimeCollaboration_etal_2021} and it is unclear in which direction a correlation would trend \citep[see e.g. discussion in][]{James_etal_2021_1} and we do not expect a significant effect. We therefore ignore such effects and assume a constant width distribution. In order to generalise the formulation of sensitivity to other telescopes, we utilise a general form thus

\begin{equation}
    \eta(\mathrm{\mathrm{DM}},w) =
    \dfrac{\eta_{0}}{\sqrt{\left(\frac{8.3 \mathrm{DM} \Delta \nu }{10^3 \nu_{c}^{3} w}\right)^2 + \left(\frac{\Delta t}{w}\right)^2+ 1}}\text{,}
 \label{eqn:EtaDM}
\end{equation}
\\

\noindent where $\nu_{c}$ is the observing centre frequency and $\eta(\mathrm{DM},w)$, $\Delta\nu$ and $\Delta t$ are the instrument's back-end sensitivity, bandwidth and time resolution respectively.

The FRB pulse-width is assumed to follow a log normal distribution with mean sensitivity given by eq.(\ref{eqn:DMEfficiency}), where the maximum burst search width, $w_m$, is taken to be $32\,\mathrm{ms}$ \citep{Arcus_etal_2020}\footnote{The maximum burst search width will change with the detection algorithm used, nonetheless, it is not expected to have a significant impact on our results since the sensitivity beyond 32 ms is negligible and the number of bursts of comparable width would be relatively small. We therefore assume $w_m=32$ ms for both telescopes.}.

\begin{equation}
    \begin{split}
    \bar{\eta}({\mathrm{DM}}) = \dfrac{1}{\sqrt{2 \pi} \ln \sigma} \int_{0}^{w_{m}} & w^{-\frac{3}{2}}
    \eta(\mathrm{DM},w) \\ & e^{- (\ln w - \ln \mu)^{2} / (2 \ln^{2} \sigma)} dw \mathrm{,}
    \end{split}
    \label{eqn:DMEfficiency}
\end{equation}
\\

\noindent and the burst-width mean and standard deviation are given, respectively, by $\mu$ and $\sigma$.

\begin{figure}
    \centering
    \includegraphics[scale=\PlotSize]{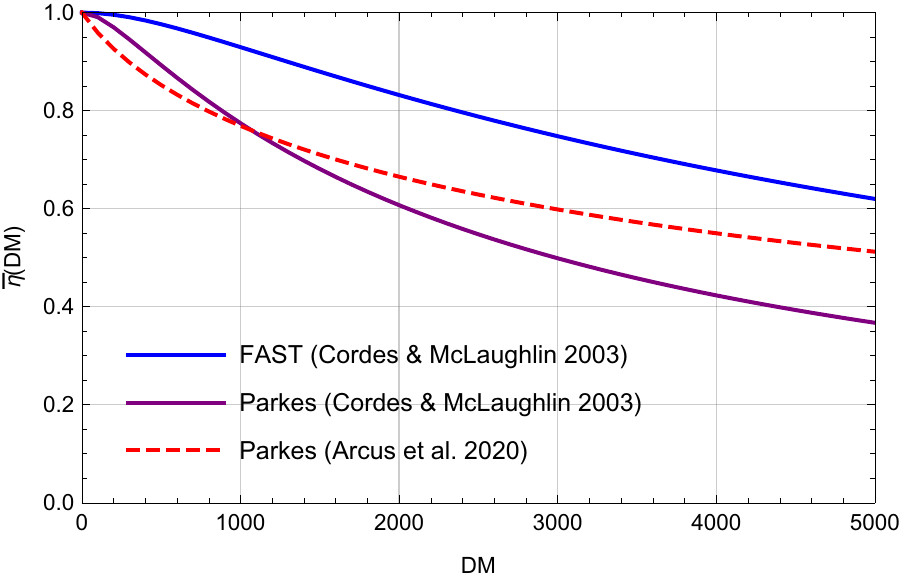}
    \caption{The normalised mean efficiency for Parkes and FAST using eq.(\ref{eqn:DMEfficiency}) \citep{Cordes_McLaughlin_2003} and the telescope parameters of Table \ref{tab:TelescopeParams}. For reference, we include the normalised mean efficiency curve attained for Parkes, using the best-fit method detailed in \citet{Arcus_etal_2020}.
    }
    \label{fig:NormalisedMeanEfficiency}
\end{figure}

\subsubsection{Beam Pattern}
In order to incorporate the beam pattern into the model, we first assess the analytic forms for the Airy relative to the Gaussian beam functions for Parkes, given the similarities in the antenna feeds between the two telescopes. The Airy and Gaussian beam functions utilised are given by eq.(\ref{eqn:AiryBeamFunction}) and eq.(\ref{eqn:GaussianBeamFunction}) respectively

\begin{equation}
    B(\theta) = \left(\dfrac{2 {\mathrm{J}}_{1}(k a \sin \theta)}{k a \sin \theta}\right)^{2}
\label{eqn:AiryBeamFunction}
\end{equation}
\\

\noindent and

\begin{equation}
    B(\theta) = A \exp \left[ - \frac{\theta^2}{2 \theta_b'^2} \right] + (1-A) \left[ - \frac{\theta^2}{2 \theta_r'^2} \right]
\label{eqn:GaussianBeamFunction}
\end{equation}
\\

\noindent where $\mathrm{J}_{1}(x)$ is the Bessel function of the first kind of order 1, $k$ the wave number at the receiver centre frequency, $\nu_{c}$, $a$ the Airy disc radius and $\theta$ the off bore-sight angle to the source.

We utilise the Gaussian formulation of \citet[][see Figure 5 and eq.(8) therein]{Macquart_Ekers_2018_1}, where $1-A=0.0015$, $\theta'_{b}=(14'/3)/\sqrt{2 \ln{2}}$ and $\theta'_{r}=(1.5^{\circ}/2)/\sqrt{2 \ln{2}}$. The additional term of the Gaussian beam function is introduced for Parkes in order to account for the off-axis response that extends scattering out to approximately $1.5^{\circ}$. This effect is due to the pedestal and its side-lobes, resulting from scattering from the base of the $6\mathrm{m}$ focus cabin, which is not included in the Airy beam \citep{Macquart_Ekers_2018_1}.

Given FAST will not incur the same pedestal and cabin scattering effects, since the dish is much larger than similarly sized antenna feed blockage at the focus, and that the DM distribution is generally insensitive to the form of the beam function \citep{James_etal_2021_1}, we choose to use the Airy function in our modelling.

\subsubsection{Burst Energy Distribution}
We use a reference source of 10\funits at $z=1$ with fluence spectral index $\alpha=0$ throughout our modelling. Furthermore, we assume an energy function described by a power law given by eq.(\ref{eqn:EnergyCurveDistribution}) \citep{Macquart_Ekers_2018_2,Arcus_etal_2020}

\begin{equation}
    \theta_{E}(E_{\nu};\gamma) = \dfrac{\gamma-1} {E^{1-\gamma}_{\mathrm{min}} - E^{1-\gamma}_{\mathrm{max}}} E^{-\gamma}_{\nu}\mathrm{,}
    \label{eqn:EnergyCurveDistribution}
\end{equation}
\\

\noindent where $\theta_{E}$ is the event rate energy function.

We choose an energy function spanning seven decades, i.e., $\sim 1.28 \times \left[10^{22}, 10^{29}\right]$ \JperHz, encompassing the upper region found by \citet[][see their Figure 2 wherein an absence of sources above $\sim10^{27}$\JperHz is noted]{Shannon_etal_2018}.

The minimum, $F_{\mathrm{min}}$, and maximum, $F_{\mathrm{max}}$, fluences corresponding to $E_{\mathrm{min}}$ and $E_{\mathrm{max}}$ respectively, are related via eq.(\ref{eqn:EnergyFluenceRelation})

\begin{equation}
    E_{\nu}(F_{\nu},z;\alpha) = \dfrac{4 \pi D_{L}^{2}(z)}{(1+z)^{2-\alpha}}F_{\nu}\mathrm{,}
    \label{eqn:EnergyFluenceRelation}
\end{equation}
\\

\noindent where $D_{L}(z)$ is the luminosity distance at redshift, $z$.

\citet{Li_etal_2021} have reported a bimodal energy distribution inferred from multiple pulses of the repeater \FRB{121102} using FAST. The extent to which the energy distribution of this particular object reflects the energy distribution of the entire population is unclear. We note that at high energies the energy distribution of \FRB{121102} is similar to that of a power-law, hence we retain the power-law in our modelling.

We initially conduct a DM distribution re-fit for Parkes and ASKAP events using the original FRB event data and model of \PaperI, treating \EMax as a free parameter. The fitted value for \EMax was found to differ only marginally to that assumed in the original work and consistent with both \citet{Luo_etal_2020} and \citet{James_etal_2018}. We therefore retain the same values of \EMin and \EMax per \PaperI\: (see Table \ref{tab:FinalParams}). Furthermore, we conduct a preliminary investigation into the effect of utilising a power law versus a Schechter function \citep[see, e.g.,][]{Lu_Piro_2019, Luo_etal_2020} for the energy function. We examine the effect that the different functional forms have on the DM distributions over the chosen energy range and we find no discernible difference between the resultant DM distributions: a result consistent with \citet{James_etal_2021_1}. We therefore choose to retain the power law formulation described by eq.(\ref{eqn:EnergyCurveDistribution}) in our modelling.

\subsection{DM Distribution Modelling}
To model the FAST DM distribution, we utilise the FRB population parameter-set determined from \PaperI\: and listed in Table \ref{tab:FinalParams}, as determined from simultaneous fits to Parkes and ASKAP data. We do so on the basis of the similarities between the FAST and Parkes telescope back-ends and until an increased sample size, hence improved statistics, become available.

We utilise a pulse-width mean of $\mu=3.44\;\mathrm{ms}$ and standard deviation of $\sigma=3.44\;\mathrm{ms}$ \citep{Arcus_etal_2020}. Using these parameters, in conjunction with the telescope parameters of Table \ref{tab:TelescopeParams}, we model the DM distribution via eq.(\ref{eqn:DMDistBeamCorrected})

The normalised mean efficiency for Parkes and FAST, via eq.(\ref{eqn:DMEfficiency}), is depicted in Figure \ref{fig:NormalisedMeanEfficiency} along with the normalised mean efficiency for Parkes, using the best-fit method of \PaperI. We note that the true response of an FRB detection algorithm is complex and may only be fully determined via FRB injection tests, such as those performed by UTMOST \citep{Gupta_etal_2021}. Deficiencies in this formulation are unlikely to have significant consequences for our results since the differences in the low DM range differ by $\lesssim 10\%$ up to $\textrm{DM} \sim2000$ \dmunits -- see Figure \ref{fig:NormalisedMeanEfficiency}. This may become increasingly significant at large DM, however, where the analytic formulation may need to be adjusted.

Integration over the FoV is performed by setting the beam integration limit $\theta_{b} = 2\cdot\theta_{\text{FWHM}}$, the point of beam overlap given the receiver feed spacing for both telescopes \citep{Staveley-Smith_etal_1996,Dunning_etal_2017}.

\begin{table*}
\caption{DM distribution model parameters used herein. The FRB population parameters are taken, \textit{a priori},  from \citet{Arcus_etal_2020}. The energy function slope, $\gamma$, represents the 68\% confidence interval and we adopt the fixed constraint of $\alpha=1.5$ \citep{Macquart_etal_2019}.
}
\renewcommand{\arraystretch}{1.3}
\label{tab:FinalParams}
\begin{threeparttable}[b]
\begin{tabular}{cccccl}
    \midrule
    $n$ & $\alpha$\tnote{$\dagger$} & $\hat{\gamma}$     & \EMin\tnote{$\dagger$} & \EMax\tnote{$\dagger$} & Evolution  \\
        &                             &                  & \multicolumn{2}{c}{(\JperHz)}                    & Model      \\
    \midrule
    0   & 1.5\tnote{$\ddagger$}     & 1.5\uclu{0.2}{0.2} & $1.28 \times 10^{22}$  & $1.28 \times 10^{29}$  & None       \\
    1   & 1.5\tnote{$\ddagger$}     & 1.8\uclu{0.1}{0.1} & $1.28 \times 10^{22}$  & $1.28 \times 10^{29}$  & Linear     \\
    2   & 1.5\tnote{$\ddagger$}     & 2.2\uclu{0.1}{0.1} & $1.28 \times 10^{22}$  & $1.28 \times 10^{29}$  & Quadratic  \\
    \midrule
\end{tabular}
\begin{tablenotes}
\item [$\dagger$] A set constraint.
\end{tablenotes}
\end{threeparttable}
\end{table*}

\begin{table*}
\caption{
The telescope parameters used to model the DM Distributions for Parkes and FAST. References are: (1) \citet{Staveley-Smith_etal_1996}; (2) \citet{Dunning_etal_2017}; (3) \citet{Li_etal_2018}; (4) \citet{Keane_etal_2018}; (5) \citet{Connor_2019}; (6) \citet{Luo_etal_2020}; \& (7) \citet{Gardenier_vanLeeuween_2020}.}
\begin{threeparttable}[b]
\begin{tabular}{lclll|ll}
    \hline
    Description                     & Symbol                & Units         & \multicolumn{2}{c}{FAST}      & \multicolumn{2}{c}{Parkes}\\
                                    &                       &               & Parameter &Reference          & Parameter &Reference\\
    \hline
    Frequency resolution            & $\Delta \nu$          & MHz           & 0.122     & (3)               & 0.39      & (5)\\
    Time resolution                 & $\Delta t$            & ms            & 0.196608  & (7)               & 0.064     & (5)\\
    Observing centre frequency      & $\nu_{c}$             & GHz           & 1.250     & (7)               & 1.352     & (7)\\
    FWHM beam-width                 & $\theta_{b}$          & $^\circ$      & 0.047     & (2)               & 0.204     & (1)\\
    Fluence limit (radiometer)      & $F_{0}$               & \funits       & 0.0146    & (6)               & 0.5       & (4)\\
    \hline
\end{tabular}
 \end{threeparttable}
\label{tab:TelescopeParams}
\end{table*}

\begin{table*}
\caption{The published FAST FRB events used in our analysis.}
\begin{tabular}{lcccll}
    \hline
    Designation & $\mathrm{DM}_{\mathrm{Obs}}$          & $\mathrm{DM}_{\mathrm{MW}}$  & DM Model  & Reference\\
                & \multicolumn{2}{c}{(pc\,cm$^{-3}$)}   &                                           &\\
    \hline
    \FRB{181123} & 1812.0 & 149.5 & YMW16  & \citet{Zhu_etal_2020} \\
    \FRB{181017} & 1845.2 & 34.6  & NE2001 & \citet{Niu_etal_2021} \\
    \FRB{181118} & 1187.7 & 71.5  & NE2001 & \citet{Niu_etal_2021} \\
    \FRB{181130} & 1705.5 & 38.2  & NE2001 & \citet{Niu_etal_2021} \\
    \hline
\end{tabular}
\label{tab:FASTFRBs}
\end{table*}

\begin{figure*}
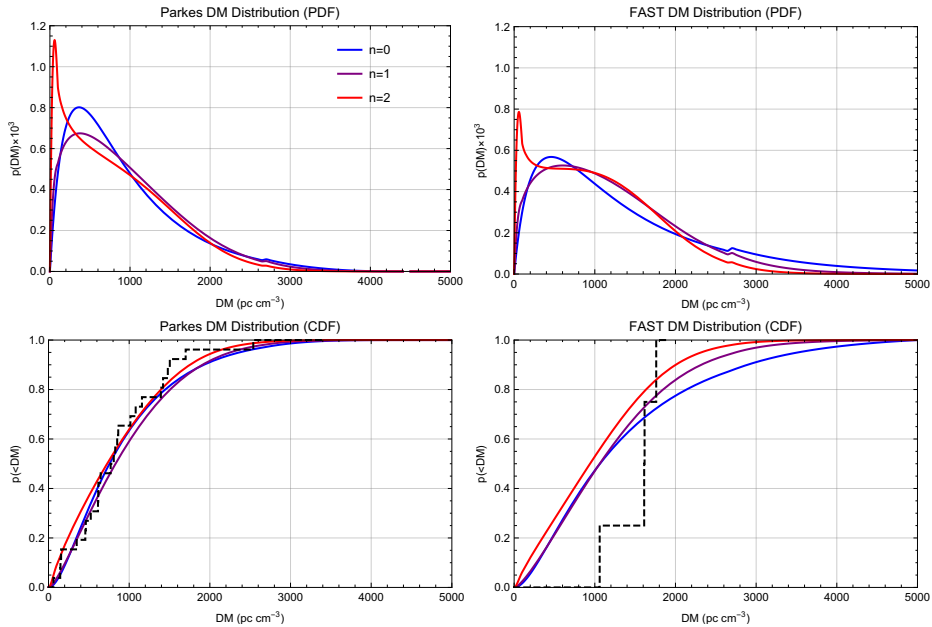

\begin{tabular}{iii}
    Graphics/ParkesPDF & Graphics/FASTPDF \\
    Graphics/ParkesCDF & Graphics/FASTCDF \\
\end{tabular}
\caption{The Parkes (left panels) and FAST (right panels) DM distributions as PDFs (top panels) and CDFs (bottom panels). The curve colours relate to redshift evolutionary scenarios: blue, $n=0$, none; purple, $n=1$, linear; and red, $n=2$, quadratic evolution with respect to the CSFR. The Parkes curves were determined by simultaneously fitting multiple population parameters for Parkes and ASKAP data in \citet[][]{Arcus_etal_2020}, whereas the FAST distributions are predicted using these same fitted population parameters of Table \ref{tab:FinalParams}. The CDFs are derived from the FRB observations from Parkes and FAST events in the lower panels (black dashed curves). FAST data are listed in Table \ref{tab:FASTFRBs}, whilst Parkes event data are drawn from \citet[][Table 1]{Arcus_etal_2020}. A Kolmogorov-Smirnov (K-S) test of the FAST event data, with respect to the computed DM distribution for the three evolutionary models, yields p-values of: $p'(n=0)=0.22$; $p'(n=1)=0.19$; and $p'(n=2)=0.12$.}
\label{fig:ParkesFASTDistributions}
\end{figure*}

\section{Discussion}
\label{sec:Discussion}
The Parkes and FAST DM distributions are depicted in Figure \ref{fig:ParkesFASTDistributions}, as both Probability Density Functions (PDFs) and Cumulative Distribution Functions (CDFs), along with the corresponding CDFs from observed FRB events. The FAST sample is listed in Table \ref{tab:FASTFRBs} and the Parkes sample (depicted for reference purposes) is drawn from \citet[][Table 1]{Arcus_etal_2020}.


Parkes DM distribution curves of Figure \ref{fig:ParkesFASTDistributions} were determined by simultaneously fitting multiple population parameters between Parkes and ASKAP as undertaken in \PaperI. The three evolutionary models for Parkes in Figure \ref{fig:ParkesFASTDistributions} are therefore good fits as they use fitted variables. The modelled FAST distributions are, however, predicted, using the same fitted population parameters of Table \ref{tab:FinalParams} and the FAST telescope parameters of Table \ref{tab:TelescopeParams}. The modelled DM distributions (solid curves) show a similar overall shape and suggest that FAST's higher sensitivity, compared to other FRB-finding instruments, gives it a superior ability to discriminate between redshift evolutionary models at DM $\gtrsim1400$\dmunits.

A Kolmogorov-Smirnov (K-S) test of the FAST observational data, with respect to the modelled DM distributions for the three evolutionary models, yields p-values of: $p'(n=0)=0.22$; $p'(n=1)=0.19$; and $p'(n=2)=0.12$. This indicates there is only mild evidence that the observations are inconsistent with all of the evolutionary models, notwithstanding statistical robustness due to the limited sample size.

The observed FAST DM distribution suggests an apparent paucity of FRB events at low DM (i.e., $\mathrm{DM}\lesssim 1000\text{ pc cm}^{-3}$) in relation to the modelled scenarios, along with a potential sharp rise in the CDF between $1000\text{ pc cm}^{-3}\lesssim\mathrm{DM}\lesssim2000\text{ pc cm}^{-3}$. None of the evolutionary models predict a paucity of events at low DM, however. This could conceivably be due to a number of factors, including:
\begin{enumerate}
    \item A complicated energy distribution or break in a single power law distribution function (e.g. a turn-over to a flatter power-law at low energies would reduce the number of near-Universe FRBs).

    \item Searches may be biased against either low or high DM events (e.g., a deficit of low-DM FRBs may result from  a bias against low-DM events due to RFI discrimination, -- see, e.g., \citet[][]{ChimeCollaboration_etal_2021} regarding bias against low DM events due to RFI).

    \item The area of sky surveyed with a high-sensitivity telescope may be sufficiently small such that the distribution of FRBs on the sky is non-uniform (i.e., the DM distribution is inhomogeneous due to cosmic variance; such an effect will likely remain untestable until sufficient deep surveys over large areas become available).

    \item The FRB sample may contain outliers (e.g., \FRB{20190520B} has recently been determined to have a large \DMHOST excess, $\mathrm{DM}_{\mathrm{Host}}=903^{+72}_{-111}\mathrm{\,pc\,cm^{-3}}$ \citet{2021arXiv211007418N}), which may explain the bulk of the paucity observed given the low number statistics involved\footnote{Presently we do not account for such outliers, neither in modelling nor data vetting, so as to not bias our analysis.}).
\end{enumerate}

Event rates are more likely to be affected by evolution at higher look-back times -- i.e., at higher redshifts or DMs. We do not consider the low DM paucity to be attributable to evolution in the sensitivity regime of FAST. This raises the interesting prospect of whether a break in the energy function exists. Further samples will be required to break the degeneracy between the evolution models and the energy curve slope: local FRBs with fluence and redshift measurements would be required.

Whilst the CHIME collaboration recently published its first FRB catalogue \citep{ChimeCollaboration_etal_2021}, we note that the CHIME telescope has a very different frequency range to FAST and it is unclear whether these results would be scalable for a comparative analysis. Furthermore, CHIME has a significantly different and complicated beam-pattern that would need to be treated in such an analysis. A new factor, however, is that CHIME provides some evidence for cross correlation between FRBs with large scale cosmic structure \citep{Rafiei_etal_2021}: a correlation between galaxies at lower redshift (i.e., $ 0.3 \lesssim z \lesssim 0.5$). Additionally, a correlation of FRBs at higher $\textrm{DM}\sim800$ \dmunits are noted, which may indicate that FRBs are arriving with DMs higher than expected, potentially resulting in the observed FAST FRB paucity at low-DM and an excess of FRBs with higher DM.

\citet{Macquart_Ekers_2018_2} discussed at length the potential to detect the helium ionisation signal of the IGM in the redshift and DM distributions using the \ME model. We note that this signal is visible in the modelled FAST DM distribution of Figure \ref{fig:ParkesFASTDistributions} (top right panel), at $\textrm{DM} \sim 2660$\dmunits (assumed to occur at $z=2.5$). We therefore expect FAST to be well placed to probe this signature should sufficient FRB events be detected spanning this region of DM-space.

The DM distributions modelled herein do not account for evolution of $\mathrm{DM}_{\mathrm{Host}}$ with redshift. The contributions to the DM of the IGM and host galaxy are inseparable and the problem is unconstrained. The host contribution is often treated as a non-redshift evolving parcel of plasma via the term $\mathrm{DM}_{\mathrm{Host}}/(1+z)$, where \DMHOST is interpreted as dispersion in the host galaxy's rest frame \citep[see, e.g., ][]{Macquart_etal_2020}. A detailed investigation of this aspect is out of scope for this paper, and we defer such work to a future paper, which may be explored via simulation.

\section{Conclusion}
\label{sec:Conclusion}
We extend and apply the \ME model along with the fitted multiple population parameters determined in \PaperI\, to the published FAST FRB events in order to compare the predicted DM distributions of FAST with respect to Parkes. We model three redshift evolutionary scenarios and assume a log power law energy function.

Whilst the published FAST sample size is limited, we observe a paucity of events at low DM relative to all evolutionary models, including a sharp rise in the observed CDF in the region of $1000\lesssim\mathrm{DM}\lesssim2000$\,pc\,cm$^{-3}$. (These two traits are not separate effects.) These observations are consistent with statistical fluctuations ($p'(n=0)=0.22$; $p'(n=1)=0.19$; and $p'(n=2)=0.12$). Whilst these traits could be due to a number of factors including statistical fluctuations ($p \le 0.22$) due to the small sample size, it may point towards a break in the energy function at the low energy regime or to an energy function turn-over.

The energy distribution is unlikely to be able to be adequately determined until further events are detected in this high-sensitivity regime. To break the degeneracy between the energy distribution and redshift evolution, we would require FRBs with similar energy at different redshifts. Identifying rare high-energy FRBs at lower redshift will require large FoV survey telescopes.

FAST has the greatest sensitivity to FRB source evolution in the region $2000 \lesssim \mathrm{DM} \lesssim 3000$ \,pc\,cm$^{-3}$, as indicated by the relative separation of the CDFs for the different evolutionary models in Figure \ref{fig:ParkesFASTDistributions}, suggesting this would be the best DM region to constrain evolution.

 The cooled Parkes Phased Array Feed, currently under construction, will provide up to 88 beams; increasing the FoV by a factor of $\sim5$ with similar sensitivity as the original Parkes survey and with well-characterised beams. Future instruments such as MeerKAT and later SKA-mid telescopes are expected to approach the sensitivity of FAST with potentially a much larger FoV, should sufficient coherent beams be formed. These instruments should increase the high-redshift FRB sample size thus improving the statistics.


\clearpage
\section*{Acknowledgements}
WRA acknowledges the contribution of an Australian Government Research Training Program Scholarship in support of this research. CWJ acknowledges support by the Australian Government through the Australian Research Council's Discovery Projects funding scheme (project DP210102103).

\section*{Data availability}
Data underlying this article are available within this article.



\bibliographystyle{mnras}
\bibliography{References.bib} 




\appendix
\section{Model Symbols \& FRB Population Parameters}
\subsection{Model Symbols \& Fitted Results}
\begin{table*}
\caption{Symbol definitions relevant to the \ME model taken from \citet{Arcus_etal_2020} and provided here for convenience.}
\begin{tabular}{ll}
    \hline
    Symbol                      & Definition \\
    \hline
    $G$                         & Gravitational constant \\
    $m_{p}$                     & Proton rest mass \\
    $z$                         & Redshift \\
    $c$                         & Speed of light \textit{in vacuo} \\
    $H_{0}$                     & Hubble constant at the present epoch \\
    $E(z)$                      & Dimensionless Hubble parameter $E(z)=\sqrt{\Omega_{m} (1+z)^{3} + \Omega_{k} (1+z)^{2} + \Omega_{\Lambda}}$ \\
    $H(z)$                      & Hubble constant at an arbitrary redshift $z : H(z) =H_{0} E(z)$ \\
    $D_{H}$                     & Hubble distance \\
    $D_{M}$                     & Comoving distance \\
    $D_{L}$                     & Luminosity distance \\
    $R_{F}$                     & Total (fluence) differential FRB event rate in the observer's frame \\
    $\Omega_m$                  & Matter density (baryonic and dark) \\
    $\Omega_{\Lambda}$          & Vacuum density \\
    $\Omega_k$                  & Spatial curvature density \\
    $\Omega_b$                  & Baryonic matter density \\
    $\alpha$                    & Fluence spectral index defined such that $F_{\nu} \propto \nu^{-\alpha}$ \\
    $\gamma$                    & Energy power-law index \\
    $F_{0}$                     & Fluence survey limit at $\mathrm{DM}=0$ \\
    $F_{\mathrm{0,P}}$          & Fluence survey limit of the Parkes telescope at $\mathrm{DM}=0$ \\
    $F_{\mathrm{0,A}}$          & Fluence survey limit of the ASKAP telescope at $\mathrm{DM}=0$ \\
    $F_{\nu}$                   & Fluence (energy spectral density per unit area) \\
    $F_{\mathrm{min}}$          & Minimum fluence for luminosity function \\
    $F_{\mathrm{max}}$          & Maximum fluence for luminosity function \\
    $E_{\nu}$                   & Spectral energy density \\
    $E_{\mathrm{min}}$          & Lower spectral energy density bound for the event rate energy function \\
    $E_{\mathrm{max}}$          & Upper spectral energy density bound for the event rate energy function \\
    $\displaystyle dR_{F}/dz$   & Fluence-based redshift distribution \\
    $\displaystyle dR_{F}/d\mathrm{DM}$  & Fluence-based DM distribution \\
    $\overline{\mathrm{DM}}(z)$          & Mean DM for the homogeneous IGM \\
    $X_{\mathrm{e,H}}$          & Fraction of ionised Hydrogen in the homogeneous IGM \\
    $X_{\mathrm{e,He}}$         & Fraction of ionised Helium in the homogeneous IGM \\
    $\psi_{n}(z)$               & Event rate per comoving volume as a function of redshift: $\psi_{n}(z) \propto \Psi^{n}(z)$ \\
    $\Psi(z)$                   & The cosmic star formation rate (CSFR) per comoving volume \\
    $n$                         & Exponent of the redshift evolutionary term per comoving volume \\
    \hline
\end{tabular}
\label{tab:ListOfSymbols}
\end{table*}

\clearpage
\bsp    
\label{lastpage}
\end{document}